\documentclass[aps,twocolumn,prl,superscriptaddress,amsmath,showpacs,tightenlines]{revtex4}
\usepackage{graphicx}
\usepackage{epsfig}

\begin{document}
%
%
\title{Efficient quantum circuits for one-way quantum computing}
%
%
\author{Tetsufumi Tanamoto}
\affiliation{Corporate R \& D center, Toshiba Corporation,
Saiwai-ku, Kawasaki 212-8582, Japan}

\author{Yu-xi Liu}
\affiliation{Frontier Research System, The Institute of Physical
and Chemical Research (RIKEN), Wako-shi, Saitama 351-0198,
Japan}
\affiliation{CREST, Japan Science and Technology Agency (JST), 
Kawaguchi, Saitama 332-0012, Japan}


\author{Xuedong Hu}
\affiliation{Department of Physics, University at Buffalo, SUNY,
Buffalo, New York 14260-1500,USA}

\author{Franco Nori}
\affiliation{Frontier Research System, The Institute of Physical
and Chemical Research (RIKEN), Wako-shi, Saitama 351-0198,
Japan}
\affiliation{CREST, Japan Science and Technology Agency (JST), 
Kawaguchi, Saitama 332-0012, Japan}
\affiliation{Physics Department, MCTP, The University of
Michigan, Ann Arbor, Michigan 48109-1040, USA}

\date{\today}

\begin{abstract}
While Ising-type interactions are ideal for implementing controlled phase flip
gates in one-way quantum computing, natural interactions between solid-state
qubits are most often described by either the XY or the Heisenberg models. 
We show an efficient way of generating cluster states directly using either
the iSWAP gate for the XY model, or the $\sqrt{\rm SWAP}$ gate for the
Heisenberg model.  Our approach thus makes one-way quantum computing more
feasible for solid-state devices.
\end{abstract}
\maketitle

%
One- and two-qubit gate operations are essential ingredients for quantum
information processing.  Significant theoretical and experimental efforts have
been devoted to study how qubits interact with external perturbations and
between themselves~(see,
e.g.,~\cite{Yamamoto,You,Imamoglu,Loss,Petta,Kane}).  Many solid-state
qubits have inter-qubit interactions described by various kinds of exchange
Hamiltonians: XY~\cite{You,Imamoglu}, XXZ, or the isotropic Heisenberg
exchange model~\cite{Loss,Petta,Kane}, {\it rather than} the Ising model
\cite{Yamamoto}~(e.g., Table 1).  
For Ising interactions, two-qubit gates such as the controlled NOT~(CNOT) and 
controlled phase flip~(CPF) are obtained
by turning-on the spin-spin interaction just {\it once}.
For non-Ising interactions, these two-qubit
gates are more difficult to implement: both CNOT and CPF gates require 
turning-on the two-qubit interaction {\it at least twice},
in addition to several single-qubit gates ~\cite{Schuch,Burkard}.
%
%

In recent years there has been steady progress towards the tunable coupling of
flux \cite{coupling} and spin qubits \cite{Petta,Wu}.  However, in general, 
qubit interactions are still difficult to control precisely.  Furthermore,
turning-on inter-qubit interactions can open new decoherence channels.  
For instance, the Heisenberg exchange interaction between electrons is
electrostatic in nature.  Turning it on makes the spin system vulnerable to
charge fluctuations in the environment~\cite{Hu}.  In order to improve the
reliability of a solid state quantum circuit, it is generally desirable to
have as few inter-qubit interaction operations as possible.  In other words,
universal quantum gates should minimize the number of two-qubit operations.

%
%
One-way quantum computing is a novel measurement-based 
approach~\cite{Briegel,Nielsen,Walther}, which starts with the creation of 
a highly entangled cluster state using CPF gates for Ising interactions.
However, for the XY, XXZ, and the
Heisenberg exchange interactions, each CNOT or CPF gate consists of at least
{\it two} iSWAP or $\sqrt{\rm SWAP}$ gates 
(as opposed to one for Ising interactions).  Here we tailor one-way
quantum computing to the inter-qubit interactions actually present in
solid
state nanostructures.  In particular, we show that the relatively cumbersome 
and expensive CPF and CNOT gates can be replaced by a {\it single-application} of an
iSWAP~(XY model), $\sqrt{\rm SWAP}$~(Heisenberg model), or generalized 
$\sqrt{\rm SWAP}$
gate (XXZ model), without additional overheads.  This change of the underlying
two-qubit gate in the quantum circuit allows much simpler, faster, and more
robust computations.
In addition, we demonstrate that iSWAP gates are particularly useful in the
construction of large cluster states.
We also show that a measurement, combined with either the XY or the XXZ
interaction, can further improve gate efficiencies in solid state quantum 
computation.
%
%
Our approach is reminiscent of recent theoretical studies in photonic 
qubits~\cite{Rudolph}, where polarization beam splitters, with 
postselection by photon detection (i.e., measurement), have been shown to 
generate cluster states efficiently without using true CNOT gates.
{\it $(J,J^z;t)$-gate.}--- The XY, XXZ, and Heisenberg models are described by
the Hamiltonian  
$
H=\sum_{i<j}H^{(ij)}
$
with 
\begin{equation}
H^{(ij)}= J_{ij}(
\sigma_{i}^{x}\sigma_{j}^{x}+
\sigma_{i}^{y}\sigma_{j}^{y})
+J_{ij}^z\;\sigma_{i}^{z}\sigma_{j}^{z},
\end{equation}
where $\sigma_i^\alpha$ ($\alpha=x,y,z$) are the Pauli matrices acting on the
$i$-th qubit with qubit basis $|0\rangle =|\!\!\downarrow\rangle$ and
$|1\!\rangle = {|\!\uparrow\rangle}$.  For simplicity we take $J=J_{ij}$ and
$J^z=J^z_{ij}$.  The XY model then corresponds to $J^z=0$, and the Heisenberg
model to $J^z=J$.  In the case of two qubits, $H^{(12)}= J (
\sigma_{1}^{x} \sigma_{2}^{x} + \sigma_{1}^{y} \sigma_{2}^{y} )
+ J^z~\sigma_{1}^{z} \sigma_{2}^{z}$ leads to a two-qubit evolution 
described by $U^{(12)}(t)=e^{-itH^{(12)}}$ ($\hbar = 1$), so that 
\begin{equation}
|01\rangle \rightarrow A|01\rangle + iB|10\rangle, \   
|10\rangle \rightarrow A|10\rangle + iB|01\rangle,
\label{eqn:AB}
\end{equation} 
with $A\equiv e^{-2iJ^zt}\cos 2Jt$ and $B\equiv e^{-2iJ^zt}\sin 2Jt$, while
$|00\rangle$ and $|11\rangle$ are unchanged (an overall phase factor
$e^{iJ^zt}$ has been omitted).  Hereafter, we call this very general
operation of turning-on $H^{(12)}$ for a time period $t$, the
$(J,J^z;t)$-gate.  The iSWAP gate is obtained when $J^z=0$ and $t = \tau_{\rm
iswap}=\pi/(4J)$, and the $\sqrt{\rm SWAP}$ gate is obtained when $J=J^z$ and
$t = \tau_{\sqrt{\rm swap}} = \pi/(8J)$.
%
%
The conventional CNOT or CPF gate requires two iSWAP gates 
for the XY model~\cite{Schuch} or two $\sqrt{\rm SWAP}$ gates 
for the Heisenberg model~\cite{Loss,Schuch,Burkard},
%
%
%
%
%
plus additional single-qubit rotations. For example, 
the XY-model CNOT gate is usually described by 
$
U_{\rm cnot}^{(12)}=e^{i\frac{\pi}{4}\sigma_{1}^{z}}
e^{-i\frac{\pi}{4}\sigma_{2}^{x}}
e^{-i\frac{\pi}{4}\sigma_{2}^{z}} 
[{\rm iswap}]_{12} 
e^{-i\frac{\pi}{4}\sigma_{1}^{x}} 
[{\rm iswap}]_{12} 
e^{-i\frac{\pi}{4}\sigma_{2}^{z}} 
$ \cite{Schuch}, where $[{\rm iswap}]_{12}\equiv U^{(12)}(\tau_{\rm iswap})$.  

%
\begin{table}[t]
\begin{tabular}{lc}
\multicolumn{2}{c}{TABLE 1. Examples of inter-qubit interactions} 
\\ \hline\hline
Two-qubit interaction & Qubit system \\
\hline
Ising & charge~\cite{Yamamoto}\\
XY &   flux~\cite{You,coupling}, charge-flux~\cite{You}, phase~\cite{You}, \\
& \{charge~\cite{You}, flux~\cite{You}, spin~\cite{Imamoglu}\} in cavity\\
XXZ  &   electrons on helium~\cite{Lidar}\\
Heisenberg & spin~\cite{Loss}, donor atom~\cite{Kane}\\
\hline\hline
\end{tabular}
\end{table}

{\it Generation of cluster states using $(J,J^z;t)$-gates.}--- Cluster 
states~\cite{Briegel} are generated by a two-body evolution of the
form $S_{ij}
\equiv (1 + \sigma_{i}^{z} + \sigma_{j}^{z} - \sigma_{i}^{z} \sigma_{j}^{z})$
acting on a product state $\Pi_i|+\rangle_i$, where $|\pm \rangle_i =
(|0\rangle_i \pm |1\rangle_i )/\sqrt{2}$.  The difficulty of applying this
approach to natural non-Ising spin models is that neighboring interactions
generally do 
{\it not} commute: $[H^{(i,i-1)},H^{(i,i+1)}]\neq 0$, so that $\exp(-iHt) \neq 
\Pi_{ij} \exp[{-iH^{(ij)}t}]$.  In order to create cluster states using
these non-Ising spin interactions, {\it pairwise} bonding between qubits are
needed~\cite{Levy}.  Specifically, for a $d$-dimensional ($d$-D) qubit
array, cluster states are generated in 2$d$ steps.  First,
two-qubit cluster states are created by performing CPF operations
between pairs of nearest-neighbor qubits.  These qubit
pairs are then connected to each other via another set of CPF operations, 
and a 1-D chain
cluster state is generated.  Afterwards, two chains are connected 
resulting in a ladder structure.  Two ladder cluster states can then be
connected into 2-D cluster states, and so on. 

%
%
%
Can we further streamline this process of cluster state generation?  
An important step in optimizing a
quantum circuit for a particular type of interaction is to identify the
fastest route to a desired entanglement.  When we closely inspect 
the various spin
interactions, we find that CNOT/CPF gates are generally not the best
two-qubit gates to generate cluster states (except in the case of
Ising interactions).  Instead, a more efficient approach is to replace the CPF
gate [in the generation of pair cluster states (the first step above)] by a
single application of the $(J,J^z;t)$-gate in the general XXZ model, together
with single-qubit rotations.  The initial two-qubit state here needs to be
$(|0\rangle_1 + e^{i\theta_1}|1\rangle_1) (|0\rangle_2 + e^{i\theta_2}
|1\rangle_2)$, with $\theta_2 - \theta_1=\pi$ or $0$.  If $\theta_2 -
\theta_1 = \pi$, the duration of the $(J,J^z;t)$-gate is $t =
\pi/[4(J+J^z)]$; if $\theta_2-\theta_1=0$, $t=(\pi/4+m_s\pi/2)/(J-J^z)$,
where $m_s$ is an arbitrary integer.  After appropriate single-qubit
rotations, a two-qubit cluster state $|\Psi\rangle_{12}^C \equiv
(|0\rangle_1|+\rangle_2+|1\rangle_1|-\rangle_2)$ is generated (for
simplicity, we omit normalization coefficients).
For isotropic Heisenberg exchange interactions, where the 
$(J,J^z;t)$-gate takes the form of $\sqrt{\rm SWAP}$, we need to prepare the 
initial state $|+\rangle_1 |-\rangle_2$.  Applying $\sqrt{\rm SWAP}$ then
leads to
$
[\sqrt{\rm swap}]_{12}|+\rangle_1|-\rangle_2
= |0\rangle_1 \{|0\rangle_2 + i|1\rangle_2\}
- i|1\rangle_1 \{|0\rangle_2-i|1\rangle_2\}
$.
After two single-qubit rotations,
$\exp[i \pi(\sigma_{2}^{z}-\sigma_{1}^{z})/4]$, $|\Psi\rangle_{12}^C$
is obtained.  For XY interactions, the pulse sequence is
even simpler: A cluster state $|\Psi\rangle^{C}_{12}$ of two qubits is
simply created by applying the iSWAP gate $[{\rm iswap}]_{12}
|-\rangle_{y1}|-\rangle_{y2}$, where $ |-\rangle_{yi}\equiv (|0\rangle_i -
i|1\rangle_i)/\sqrt{2}$ is an eigenstate of $\sigma^y$.

%
%
%
Our new approach here can save more than half the time over the conventional
method during the first step of cluster state generation.  
For example, when using the two-qubit spin
Hamiltonian~\cite{Burkard} $H_s^{(ij)} = J {\vec \sigma}_i \cdot {\vec
\sigma}_j + ( {\vec B}_i \cdot {\vec \sigma}_i + {\vec B}_j \cdot {\vec
\sigma}_j)/2$, with $|{\vec B}_i|/2=J$ for simplicity, a time $t_{\rm
cpf}=\pi/J$ is needed for generating a two-qubit cluster state including
single-qubit rotations, using the conventional method.  However, using our new
method, it
takes $\tau_{\sqrt{\rm swap}} + \pi/(4J) = (3/8)(\pi/J)$, which amounts to a
$\sim$ 2.7 {\it speed-up} in time for generating a two-qubit cluster state. 
For spin qubits~\cite{Petta} based on quantum dots, with $J \sim 50$ $\mu$eV,
the time required for generating a two-qubit cluster state would be $\sim$ 15
psec.  
For a flux qubit \cite{You} in the rotating frame, the Hamiltonian is 
$\tilde{H}_{\rm fq} = H_{0} + H_{xy}$, where
$H_{0}=\sum_{i=1}^2 (\Omega^R/2) (\sigma_{i}^{x}\cos \phi_i+\sigma_{i}^{y}
\sin \phi_i)$, $H_{xy}=J (\sigma_{1}^{x} \sigma_{2}^{x} + \sigma_{1}^{y}
\sigma_{2}^{y})$, and $\Omega^R$ is the half-amplitude of the applied
classical field.  The time required to generate a two-qubit cluster state 
previously \cite{tana06} was $t_{\rm cs}^{\rm old} =
(11\pi)/(4\Omega^R) + \pi/(4J) \sim 18$ ns ($\Omega^R \sim J \sim 0.5$~GHz).
In the method proposed here, we just need $t_{\rm cs}^{\rm new} = \tau_{\rm
iswap}\sim 1.57$~ns, which is {\it over one order of magnitude faster}.

The reduction in the number of quantum gates naturally increases the
robustness of cluster state generation.  Consider a simple case where
there are phase errors in each of the one- and two-qubit gates, such that
$\theta \rightarrow \theta + \delta_\theta$ and $Jt \rightarrow Jt +
\delta_J$ respectively ($\delta_\theta,\delta_J \ll 1$).  The resulting
two-qubit state, denoted by $|\Psi\rangle_{12}^{C{\rm (error)}}$, is then
slightly
different from the target two-qubit cluster state.  The {\it fidelity} of
this state, if generated by a single $\sqrt{\rm SWAP}$ with single-qubit
rotations starting from $|+\rangle_1 |-\rangle_2$, is given by $|_{12}^{C}
\langle \Psi | \Psi \rangle_{12}^{C{\rm (error)}}|^2 \sim 1 - 2
\delta_\theta^2 -
4\delta_J^2$, which is higher than the one achieved by 
the conventional CPF gate in Refs.~\cite{Schuch,Burkard}, 
where $|_{12}^{C} \langle\Psi |
\Psi\rangle_{12}^{C{\rm (error)}} |^2 \sim 1 - 2.5 \delta_\theta^2 -
4\delta_J^2$. 
When an iSWAP gate is used, starting from $|-\rangle_{y1}|-\rangle_{y2}$, the
fidelity for two-qubit cluster state generation is $(1 + \cos 2\delta_J)/2
\sim (1 - \delta_J^2)$, which improves greatly over the previous
result~\cite{tana06} of $1 - 4\delta_J^2 - (1 - \sin^2(\pi/8))
\delta_\theta^2$. For example, the fidelity increases from 
0.84 (0.95) to 0.96 (0.99) for 20 $\% (10\%)$ errors in $\delta_\theta$ and
$\delta_J$.

\begin{figure}
\begin{center}
\includegraphics[width=6.5cm]{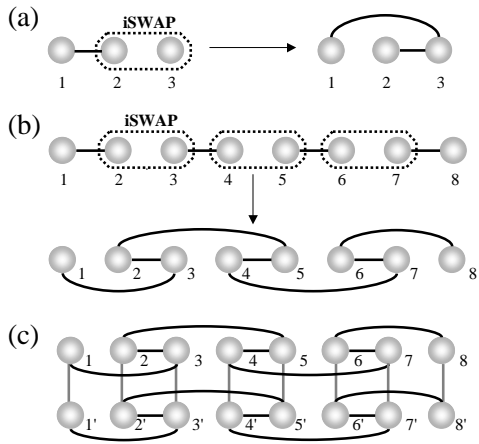}
\end{center}
\vspace{-0.7cm}
\caption{Illustration of how to generate cluster states with iSWAP gates.
Each circle represents a qubit. Each solid line represents a bond by cluster
states.
(a) A three-qubit cluster state from a two-qubit cluster state. 
(b) Creation of a chain cluster state. 
The first step is to create separated two-qubit cluster 
states. The second step is to apply an iSWAP gate 
between the 3 two-qubit cluster states. 
(c) Two chain cluster states produced in (b) are 
vertically connected by iSWAPs, thus producing a ladder cluster states.  
Afterwards, several of these can be connected to produce 
2-D cluster states. 
} \label{Fig1}
\end{figure}

{\it Generation of larger cluster states with iSWAP gates.}--- The iSWAP 
gate is not just an efficient substitute in the generation of pair
cluster states.  It can also simplify the generation of larger cluster
states.
Consider the case of generating a three-qubit cluster state.  Starting with
qubits `1' and `2' already in a cluster state, applying an iSWAP gate
between qubits `2' and `3' leads to
\begin{eqnarray}
[{\rm iswap}]_{23}|\Psi\rangle_{12}^C |+\rangle_3 
& = & |+\rangle_1 |0\rangle_3 [|0\rangle_2+i|1\rangle_2] \nonumber \\
& & + i|-\rangle_1 |1\rangle_3 [|0\rangle_2-i|1\rangle_2].
\end{eqnarray}
Additional transformations ${|0\rangle_j \rightarrow |0\rangle_j, i|1\rangle_j
\rightarrow |1\rangle_j}$ ($-\pi/2$ rotation around the $z$ axis, 
$P\equiv {\rm diag}(1,-i)$) 
for $j=2,3$ then lead to the ``twisted'' cluster 
state shown in Fig.~1(a), which is different from the conventional
three-qubit cluster state $|\Psi\rangle^{C}_{123}=|+\rangle_1 |0\rangle_2
|+\rangle_3 + |-\rangle_1 |1\rangle_2 |-\rangle_3$ by an exchange of 
the indices of qubits `2' and `3'. 
This simple example suggests that iSWAP gates 
can be used to expand cluster states even after the second step 
mentioned in the previous section.
Using the iSWAP gate with only $P$ , two-qubit cluster states can be connected
to 
make a large cluster chain as shown in Fig.~1(b). 
Moreover, cluster states in higher dimensions
can be generated in steps, as shown in Fig.~1(c).
%
%

The iSWAP gate is particularly powerful [but not the $\sqrt{\rm SWAP}$ and the
general $(J,J^z;t)$-gates] in the cluster state generation and manipulation.
This fact can be understood and illustrated by applying a $(J,J^z;t)$-gate 
between a pair of qubits that are in a pair cluster state 
$|\Psi\rangle^{C}_{12}$ and a third qubit in an arbitrary superposition
state $a_3|0\rangle_3 + b_3|1\rangle_3$, in an attempt to generate a
three-qubit cluster state $|\Psi\rangle^{C}_{123} = |+\rangle_1 |0\rangle_2
|+\rangle_3 + |-\rangle_1 |1\rangle_2 |-\rangle_3$.  To maintain the right
number of basis states in the three-qubit superposition state, we need
$A=0$ in Eq.~(\ref{eqn:AB}).  This also leaves us with $B=\pm 1$, which
allows the factorization of the three-qubit state as in the three-qubit cluster
state.  But the conditions above correspond exactly to an iSWAP gate.  A
general $(J,J^z;t)$-gate or a $\sqrt{\rm SWAP}$ gate would have created
additional terms so that additional steps are needed to clean up the state. 
In other words, the iSWAP gate combines both the power to entangle and 
the advantage of simplicity.   
Because the iSWAP gate is decomposed into a product 
of a CNOT and a SWAP gates~\cite{Schuch}, it is  
{\it fault-tolerant} as well~\cite{Gottesman}. 
Thus, we can say that cluster states in the XY model 
are constructed by {\it fault-tolerant} operations iSWAP and 
rotation $P$.

The replacement of CNOT gates by iSWAP gates has very broad 
implications in the context of the general manipulation of 
quantum information, going well beyond
one-way quantum computing and the generation of cluster states.
For example, it can be used for quantum error-corrections.
For an arbitrary qubit state, 
$|\psi\rangle=a|0\rangle+b|1\rangle$, the three-bit-flip state 
$a|000\rangle+b|111\rangle$ is conventionally obtained 
by two CNOT gates: 
$[{\rm cnot}]_{12}[{\rm cnot}]_{13}|\psi\rangle_1|0\rangle_2|0\rangle_3$.
Using iSWAP, this state can instead be generated
by $e^{i(\sigma_1^x+\sigma_2^x)\pi/4}
[{\rm iswap}]_{23}[{\rm iswap}]_{12}|\psi\rangle_1|+\rangle_2|+\rangle_3$,
which again reduces the complexity of the quantum circuit. 

In short, we have shown that we can efficiently create pair cluster states by
directly using the $(J,J^z;t)$-gates (specially the iSWAP or $\sqrt{\rm
SWAP}$ gates when the interaction is of XY or isotropic Heisenberg type)
based on naturally existing interactions.  Furthermore, all CNOT gates
can be replaced by iSWAP gates for cluster state generation and other
broader applications.

{\it Cluster fusion using iSWAP gates.}--- In general, 
Ising interaction is the most convenient
interaction for one-way quantum computing because 
it generates CPF gates directly, while the isotropic 
Heisenberg interaction is the most
cumbersome.  The XY interaction and the associate iSWAP gate 
are almost equivalent to the Ising interaction and the CNOT 
gate, so that the iSWAP can be
used to replace CNOT gates without incurring large costs in
computational resources.  
The power of the iSWAP gate can be further enhanced in one-way quantum 
computing when combined with measurements.  As an example,
we show that a large $(M+N-1)$-qubit cluster chain can be
created by joining two initially-separated $M$-qubit and $N$-qubit
cluster chains ($M$ and $N$ are arbitrary integers) using one iSWAP gate
and measurement, similar to the idea of ``qubit fusion" described in
Ref.~\cite{Rudolph}.  

\begin{figure}
\begin{center}
\includegraphics[width=4.5cm]{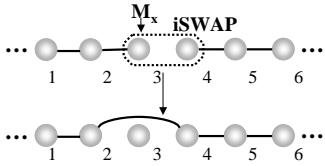}
\end{center}
\vspace{-0.75cm}
\caption{
Connecting two half-infinite cluster states 
via an iSWAP gate and measurement. After applying an iSWAP gate 
between qubit `3' and `4', 
qubit `3' is measured on the $\sigma^x$ basis following 
appropriate rotations in qubit `4' and `5'.
Qubit `3' is discarded after the measurement.} 
\label{FigBond}
\end{figure}

Consider two initially-separated qubit chains that are in cluster states, 
$|\Psi_L\rangle=\cdots S_{12}S_{23}|+\rangle_1|+\rangle_2|+\rangle_3$
and $|\Psi_R\rangle =S_{45}S_{56}|+\rangle_4$ $|+\rangle_5|+\rangle_6\cdots$. 
We connect the end of the first chain and the beginning of the second chain by
applying an iSWAP between qubits `3' and `4' (Fig.~\ref{FigBond}).  The
resulting state is 
$
[{\rm iswap}]_{34}|\Psi_L\rangle |\Psi_R\rangle =
\cdots\; S_{12}S_{56}(2|\Theta\rangle) |+\rangle_1|+\rangle_6\; \cdots
$,
where 
$|\Theta \rangle =[{\rm iswap}]_{34} |\Psi\rangle^C_{23}
|\Psi\rangle^C_{45}$.
Next we carry out a $\sigma^x$ measurement on qubit `3' (or qubit `4'), 
so that  
\begin{eqnarray}
|\Theta\rangle & \rightarrow & |+\rangle_2  |0\rangle_{4}
[(1+(-1)^{s_3}i) |0\rangle_5 +(1-(-1)^{s_3}i) |1\rangle_5 ]
\nonumber \\
&+& |-\rangle_2  |1\rangle_{4} [(i+(-1)^{s_3}) |0\rangle_5 + (i-(-1)^{s_3})
|1\rangle_5 ].
\end{eqnarray}
Here $s_3=0$ or $1$ is the result of the measurement.  After applying  
a rotation 
$e^{-i\frac{\pi}{4} \sigma_{5}^{z} }
[e^{ i\frac{\pi}{2} \sigma_{5}^{x} }]^{s_3}$, 
we obtain a $(M+N-1)$-qubit cluster state $\cdots S_{24}S_{45}| +
\rangle_2|+\rangle_4|+\rangle_5 \cdots = \cdots\{ |+\rangle_2 
|0\rangle_{4}|+\rangle_{5} + |-\rangle_2 |1\rangle_{4}|-\rangle_5 \}
\cdots$.
The fidelity of this cluster fusion, in the presence of pulse errors
$\delta_\theta$ and $\delta_J$, is 
$\sim 1 - \delta_\theta^2 (1+s_3) -2\delta_J^2$. 
Compared with a previous generation method~\cite{tana06} for flux 
qubits, the fidelity is now improved when 
$\delta_\theta \lesssim \sqrt{2}\delta_J$.

%
%
%

{\it Connection between distant qubits.}--- Last but not least, 
$(J,J^z;t)$-gates can generate a cluster state for two distant
qubits when combined with measurements.
Consider a chain of $2N$ qubits in a product of pair-wise
cluster states $\Pi_{j=1}^N |\Psi\rangle^C_{2j-1,2j}$. 
A two-qubit cluster 
state $|\Psi\rangle^C_{1,2N}$ is efficiently obtained 
as follows: 
(1) Apply $(J,J^z;t_1)$-gates  
between qubits `$l$' and `$l+1$' ($l=2,4,..$)  
that belong to neighboring two-qubit cluster states
($t_1$ is determined below); 
(2) Perform $\sigma^x$-measurements on all intermediate 
qubits $2, 3,..,2N-1$.
After these two steps, the $2N$-qubit state 
becomes 
$\{u_+|+\rangle_1 +v_-|-\rangle_1\}|0\rangle_{2N}
+\{u_-|+\rangle_1 +v_+|-\rangle_1\}
|1\rangle_{2N}$, 
with
$
\left(
\begin{array}{cc}
u_+ & u_- \\
v_- & v_+ 
\end{array}
\right)
=
\Pi_{j=1}^{N-1}\left(
\begin{array}{cc}
u_{j+} & u_{j-} \\
v_{j-} & v_{j+} 
\end{array}
\right)
$, 
$v_{j\pm}= 
 \mp(-1)^{s_{2j}+s_{2j+1}}u_{j\mp}$
and 
$u_{j\pm}=$ 
$1\pm(-1)^{s_{2j}} \exp{ 2i((-1)^{s_{2j}+s_{2j+1}}J-J^z)t_1}$, 
where $s_{2j}$ and $s_{2j+1}$ are measurement outcomes ($s_{2j},s_{2j+1}=\{0,1\}$).
The unitarity of this transformation dictates that
$\cos 2((-1)^{s_{2j}+s_{2j+1}}J-J^z)t_1=0$
(this condition is generally not satisfied 
by the uniform Heisenberg model ($J=J^z$));
(3) Finally, depending on the measurement outcome,
rotate qubit `1' appropriately, 
and we obtain $|\Psi\rangle^C_{1,2N}$.
%
%


{\it Discussions.}---
An essential ingredient of our study is to identify the exact effects of two-qubit
interactions in solid-state one-way quantum computing.  Instead of overcoming
difficulties in gate operations through
encoding in logical qubits~\cite{Lidar,Viola}, we here focus on optimizing
the gate capabilities of naturally existing interactions and build quantum
circuits accordingly.  Our approach leads to significantly simplified quantum
circuits compared to conventional ones built around CNOT/CPF gates, resulting
in faster and more robust gates.  Such simplicity should also be
useful in the fight against decoherence, especially when combined with known
approaches such as dynamic decoupling~\cite{Viola}. 
We also observe that, 
as the isotropy in the interaction decreases
from the Heisenberg model to the Ising 
model, through the XXZ and XY models, the generation of cluster states becomes 
more efficient as the required number of steps decreases.  

{\it Conclusions.}---
We have shown that $(J,J^z;t)$-gates based on known interactions in solid
state qubits (such as iSWAP gates for the XY interactions and $\sqrt{\rm
SWAP}$ gates for the isotropic Heisenberg exchange interaction) allows
significantly more efficient quantum circuits for one-way quantum computing.
The gains in circuit efficiency and robustness make solid state qubits more 
feasible.
The success of the present approach 
depends strongly on the interaction anisotropy, described by $J$ and $J^z$.
In particular, the iSWAP gate ($J^z=0$) is especially
attractive for its simplicity and its ability to entangle, so that it 
can replace the more widely used CNOT/CPF gates in a broad spectrum
of applications ranging from one-way quantum computing to
quantum error correction.

FN and XH are supported in part by the NSA, LPS, ARO, and NSF.
We thank K. Maruyama for remarks.

\vspace*{-0.2in}

\end{document}